\documentclass{elsart}
\usepackage{amsmath}

\input{tcilatex}

\begin{document}

\begin{frontmatter}

\title{Quantifying Complexity in the Minority Game }

\author{Milan Rajkovi\'c, Zoran Mihailovi\'c }

\address{Vin\v ca Institute of Nuclear Sciences 
   P.O. Box 522, 11001 Belgrade, Yugoslavia}

\begin{abstract}
 A  Lempel-Ziv complexity measure is introduced into the theory of a Minority Game (MG)  in order to  
capture some features that  volatility, one of the central quantities in this model of interacting  agents, is not able to.    
Extracted solely from the binary string of outcomes of the game complexity offers new and valuable information on collective behavior of players.  
Also, we show that an expression  for volatility may be included in the analytical expression for complexity.   
\vskip 12pt
\noindent PACS: 02.50.Le, 05.40.-a, 05.65.+b, 87.23.Ge
\end{abstract}

\begin{keyword}
Complexity measure; Minority Game; Economy; Phase transition
\end{keyword}

\end{frontmatter}

\section*{Introduction}

The Minority Game has been a topic of considerable recent attention as a
paradigm of a complex adaptive system where individal actions of each
participant (agent) give rise to cooperative effects reflected in the
collective behavior of the whole population of players. The model is also a
valuable tool for studying financial markets as particular cases of an
interacting agent systems. The central issue of this work is the complexity
of a binary string which represents outcome of the game, and its role in the
understanding of collective behavior of players. The Lempel-Ziv complexity
is used as a measure of complexity and it turns out that it captures many
features otherwise eluding usual analysis. The paper is organized such that
first a brief presentation of this complexity measure is given followed by
an introduction to the setup of the Minority Game (MG). The relationship
between phase transitions, symmetry breaking and complexity is also
presented in this section followed by a comparison with a similar work on
physical complexity in MG. Finally we discuss results and their relevance in
explaining collective behavior of players.

\section{The Complexity Measure of Lempel and Ziv}

According to the widely accepted notions of complexity [1, 2] the complexity
of a string of zeros and ones (i.e. a binary string) is given by the number
of bits of the shortest computer program which can generate the string. From
the pool of all possible programs Lempel and Ziv \cite{LempelZiv} have
chosen the ones that involve only two operations: copying and inserting, and
instead of calculating the length of the programs the calculation is focused
on a single number $c(n)$ which is a useful measure of this length. Very
briefly the algorithm consists in scanning a given $n$-digit sequence from
left to right and at each point of the sequence a test is performed to check
whether the rest of the sequence may be reconstructed by simple copying or
whether one has to insert new digits. The number of production steps
reflected in the number of newly inserted digits represents a complexity of
a binary string. Details of the algortihm may be found in the original paper
of Lempel and Ziv \cite{LempelZiv}. In order to shed more light on the
results we shall consider some analytic features of $c(n)$. The complexity
of a binary string of length $n$ in the limit of large number of
uncorrelated random digits tends to 
\begin{equation*}
\lim\limits_{n\rightarrow \infty }c(n)=c_{R}(n)\equiv n/\log _{2}n.
\end{equation*}%
$c_{R}(n)$ hence represents the asymptotic value of complexity of a binary
random string\footnote{%
We normalize all our numerically obtained complexity values by the maximal
complexity of a random string evaluated by averaging over 100 000 random
samples whose length corresponds to the length used in simulations.}.
Naturally in the case when probabilities of finding each of two digits in a
binary representation, say 1 and 0, are different from 0.5 we would expect
the complexity to be smaller than in the case when probabilites of finding
both digits are the same. In this case $c(n)$ tends to [3, 4]%
\begin{equation}
\lim\limits_{n\rightarrow \infty }c(n)=H(p)\frac{n}{\log _{2}n}=-[p\log
_{2}p+(1-p)\log _{2}(1-p)]\frac{n}{\log _{2}n},  \label{source}
\end{equation}%
where $H(p)$ is known as the source entropy. Clearly this quantity is
maximal for $p=0.5$ and once $p$ and $H(p)$ are determined it can be tested
wheather deviation of $lim_{n\rightarrow \infty }[c(n)/c_{R}(n)]$ from $1$
is due to the fact that $H(p)$ differs from $1$ (and the string is still as
random as possible) or, as is the case when $lim_{n\rightarrow \infty
}[c(n)/c_{R}(n)]<H(p),$ due to the presence of correlations in the string.

\section{Minority Game}

Minority game \cite{Challet} consists of $N$ players trying at each time
step to be in minority taking an action $a_{i}(t)=\pm 1$, where, for
example, $1$ may be interpreted as buying and $-1$ as selling an asset. The
choice is made based on the information stored as a binary time series of
the last $m$ actions taken by the minority $\mu _{t}=(\chi _{t-1},...,\chi
_{t-m})$. Each player has at his disposal the same number $s$ (but in
general different) of strategies among the $S$ possible strategies, denoted
by $s_{\pm }$, which give a prediction for the next outcome of the game
based on the history of the last $m$ outcomes. Since there are $2^{m}$
possible histories there are $S=2^{2^{m}}$possible strategies. Each strategy
is dynamically assigned a virtual payoff based on its performance and at
each time step every player uses his most successful strategy in terms of
the highest payoff (highest $-a_{i}(t)A(t)$ where the difference in the
population of agents choosing the $+$ and the $-$ sign at time $t$ is $%
A(t)=\sum\nolimits_{i}^{N}a_{i}(t)$). Therefore, this interaction rewards
the minority of agents (those who took the action $a_{i}(t)=-sign$ $A(t)$),
and hence the name minority game. If a player $i$ follows his $s$ - th
strategy and the history information is $\mu $ his action is denoted by $%
a_{s,i}^{\mu }.$ The variance $\sigma $ of $A(t)$ ($volatility$ in financial
context) represents one of the most interesting observables since the
smallest value attained by $\sigma $ represents the maximal payoff
distribution to agents and minimal global waste of resources by the
community of agents.

\section{Complexity in the Minority game}

\subsection{Phase transition, summetry breaking and complexity}

The MG undergoes a second order phase transition with symmetry breaking as
the control parameter $\alpha =2^{m}/N$ is varied [6, 10]: the system is in
the symmetric phase (temporal average of $A(t)$ conditional to $\mu (t)=\mu $
equals $0$, i.e. $\langle A^{\mu }\rangle =0$ ) for $\alpha <\alpha _{c}$
and it is in the asymmetric phase for $\alpha >\alpha _{c}$. In other words
in the symmetric phase both actions are taken by the minority with the same
frequency while in the asymmetric phase the minority prefers one action over
the other. Moreover, for $\alpha <\alpha _{c}$ all players use all their
available strategies while for $\alpha >\alpha _{c}$ a finite population of
players uses only one strategy (analogous to spontaneous magnetization in
the spin formalism). It has been suggested that behavior of $\sigma $ does
not depend on the real history of the game \cite{Cavagna}, however it was
shown that this assertion is true only in the symmetric phase \cite%
{ChalletMarsili}.\ Additionally, volatility was criticized for its
insensitivity to the real history of the game and another measure based on
the algorithmic complexity was suggested \cite{Mansilla}. Actually, in
reference \cite{Mansilla} the physical complexity (non normalized Shannon
entropy) of substrings of the string of outcomes is evaluated which is equal
to zero for random sequences, and the effect of memory size on the degree of
randomness was in the focus of the study. On the other hand the Lempel-Ziv
complexity is maximal for white noise and the focus of our work lies in
proving that this complexity measure may be used to detect phase transitions
and other information about players and the game in general that can not be
obtained by using volatility alone. From our point of view also, volatility
is inferior in many aspects to the complexity measure of Lempel and Ziv, or
variations of it, however due to its important role in theory of spin glass
systems which are analogous to the setup of MG, it would be useful if it
appears in the analytic expression for complexity. In order to clearly
illustrate the advantage of using the complexity measure of Lempel and Ziv,
we numerically simulate a game with two types of players: those who use real
history in selecting an optimal strategy and those who use assigned invented
(random) history at each time step. The behavior of $\sigma ^{2}$ is shown
in Fig. 1. Note that in this case too there is a second order phase
transition with symmetry breaking and two phases of the game can be clearly
distinguished. However, when the game is played by only one group of players
the graph of $\sigma ^{2}$ vs. memory is qualitatively the same while the
graph of complexity, presented in Fig. 2 clearly shows the difference.
Specifically, when the game is played solely by players using invented
history no phase transition is detected and complexity is practically
constant in the complete range of memory values (inset of Fig. 2). When only
players using real history take part in the game there is a phase transition
at $m=5$ indicated by the maximum value of complexity. Comparing the minimum
value of volatility with respect to memory size with the corresponding
maximum value of complexity we may expect that maximal collaboration among
the players (and minimal global waste) is achieved very close to the maximum
value of complexity. Maximal collaboration is an inherently complicated
occurrence considering the fact that none of the players are aware of
actions taken by other participants in the game. Based on these findings we
may add that history is relevant! An important information conveyed by
complexity is its ability to distinguish types of players involved in the
game based on the information contained only in the binary string of
outcomes. In the economic context this would mean that such inference may be
obtained from the direction of the sign of the order imbalance, i.e. from
returns. In addition, this finding supports the intermediate version of the
efficient market hypothesis, namely that a number of important informations,
including the ones that are not available to the players (agents) playing
the game (taking part in the financial market activity), are reflected in
the outcome (price index) at each moment of time. One more important feature
is that the maximum value of complexity (an indication of phase transition)
is \textit{independent of memory} as indicated in Fig. 3, where complexity
is presented as a function of the usual order parameter $\alpha =P/N\equiv
2^{m}/N,$ with all players relying on the true history of the game. Note
that for each maximal value of complexity the order parameter value is $\sim
0.34$, a universally accepted value at phase transition.

\subsection{Complexity and Entropy}

Inspired by the expression (\ref{source}) for source entropy, we introduce
the following entropy function 
\begin{equation}
H_{\mu }=-\frac{1}{P}\sum\nolimits_{\mu }[p^{\mu }\log p^{\mu }+(1-p^{\mu
})\log (1-p^{\mu })],  \label{entropy}
\end{equation}%
where $P$ is the state space of all histories of size $m$, i.e. $P=2^{m},$
and the summation is over the histories. We also assume that the
probabilities figuring in the above expression depend on the history of the
outcomes of the game. Namely, the probability $p^{\mu }$ represents the
probability of finding a minority sign $a_{t}=-sign$ $A_{t}$ provided a
specific history of outcomes occurred. Hence, it is a conditional probability%
\begin{equation*}
p^{\mu }=\text{ Prob \{}A(t)>0\mid \mu (t)=\mu \text{\}.}
\end{equation*}%
In the limit of large $N$, $\ A^{\mu }$ is a gaussian variable with average $%
\langle A^{\mu }\rangle $ and variance $\langle (A^{\mu })^{2}\rangle $ - $%
\langle A^{\mu }\rangle ^{2}.$ A calculation yields%
\begin{equation}
p^{\mu }=\frac{1}{2}\func{erf}c\frac{\mid \langle A^{\mu }\rangle \mid }{%
\sqrt{2(\langle (A^{\mu })^{2}\rangle -\langle A^{\mu }\rangle ^{2})}}\simeq 
\frac{1}{2}-\frac{1}{\sqrt{2\pi }}\frac{1}{\sqrt{\frac{\sigma ^{2}}{h}-1}},
\label{probability}
\end{equation}%
where $\sigma ^{2}=\langle (A^{\mu })^{2}\rangle ,$and $\langle A^{\mu
}\rangle ^{2}=h,$ and where we have expanded the erfc function to linear
order under the assumption that $\langle A^{\mu }\rangle \ll Var(A^{\mu })$.
Consequently, this entropy expression contains both $\sigma ^{2}$ and $h$
keeping the analogy with the spin glass systems (the dynamcis of the
Minority Game is similiar to spin dynamics with hamiltonian $\sigma ^{2})$ %
\cite{Challetphase}. In Fig. 4 we show the entropy given by expression (\ref%
{entropy}) for standard MG as a function of $\alpha =P/N.$ A striking
feature of this graph immediately draws attention, namely for an initial
range of the order parameter $\alpha $ the entropy is maximal $(\simeq 1)$
and then it starts decreasing at $\alpha \simeq 0.34,$ the value of the
control parameter at phase transition. In the symmetric phase of the game,
when both actions are taken by the minority with same frequency the entropy
is maximal and close to 1, and this result may be easily predicted using
expressions (\ref{entropy}) and \ (\ref{probability}). However when minority
starts preferring one action over the other the entropy starts decreasing
and a phase transition occurs. Since the Lempel Ziv algorithm is optimal in
the sense that at least for ergodic and infinite sequences complexity
approaches the Shannon entropy \cite{Ziv}, it is of interest to compare the
Shannon \ entropy and the entropy given in expression (\ref{entropy}) in
order to shed more light on the difference between the complexity shown in
Fig. 3 and the entropy presented in Fig. 4. Numerically, the Shannon entropy
from a finite binary sequence of length $N$ is usually estimated by
calculating all block (subsequence or ''word'') probabilities $%
p(s_{1},...,s_{n})$ by the standard likelihood estimate, 
\begin{equation}
\hat{p}_{n}=n_{s_{1}...s_{n}}/N  \label{p_shannon}
\end{equation}%
where n$_{s_{1}...s_{n}}$ is the number of occurrences of the block $%
s_{1},...,s_{n},$ and $s_{i}\in \{0,1\},$ and where the expression for the
Shannon entropy is 
\begin{equation}
H_{S}=\lim_{N\rightarrow \infty }-\frac{1}{N}\sum\nolimits_{\{s_{i}%
\}}p_{n}(s_{1},...,s_{n})\log p_{n}(s_{1},...,s_{n}).  \label{shannon}
\end{equation}%
Hence, the probabilities (\ref{probability}) and (\ref{p_shannon}) figuring
in the entropy expressions (\ref{entropy}) and (\ref{shannon}) respectively,
are inherently different. As a final remark we may add that in the context
of MG complexity may be linked to Nash equilibria, collaborative behavior
among players and bounded rationality \cite{Milan}.

\section{Summary}

A complexity measure based on the algorithm of Lempel and Ziv offers new
evidence that history is relevant in MG, and that maximal complexity
corresponds to the maximal collaboration among players. It reaches its
maximum at phase transition value of the order parameter regardless of the
memory size. Also, types of players involved in the game may be recognized
using the complexity measure and overall findings support the hypothesis
that important information is contained in the outcomes. Inspired by an
analytic expression for asymptotic complexity we introduce an entropy
expression which contains volatility, a central quantity in the theory of
the Minority Game, enabling links to the exact solution and analytical
results \cite{MarsChaletZecc} .

\section{Acknowledgements}

This work was partially supported by Serbian Ministry of Science and
Technology project OI 1986.

\end{document}